\documentclass{PoS}

\usepackage{cite}

\usepackage{color}
\usepackage{slashed}

\usepackage{amsmath, amsthm, amssymb}
\newtheorem{thm}{Theorem}[section]

\newtheorem{definition}[thm]{Definition}
\newtheorem{example}[thm]{Example}
\newtheorem{remark}[thm]{Remark}

\newcommand{\ep}{\varepsilon}
\newcommand{\Dx}{D_x}
\newcommand{\KK}{\mathbb{K}}
\newcommand{\QQ}{\mathbb{Q}}
\newcommand{\NN}{\mathbb{N}}
\newcommand{\ZZ}{\mathbb{Z}}

\title{{\footnotesize 
       DESY 19-225, DO-TH 19/13, SAGEX--19-35, PoS(RADCOR19)078}\\
A refined machinery to calculate large moments from coupled systems of linear differential equations}

\ShortTitle{A refined large moment machinery}

\author{Johannes Bl{\"u}mlein\\
      Deutsches Elektronen--Synchrotron, DESY,\\
      Platanenallee 6, D-15738 Zeuthen, Germany.\\
      E-mail: \email{johannes.bluemlein@desy.de}
}

\author{Peter Marquard\\
      Deutsches Elektronen--Synchrotron, DESY,\\
      Platanenallee 6, D-15738 Zeuthen, Germany.\\
      E-mail: \email{peter.marquard@desy.de}
}

\author{\speaker{Carsten Schneider} \\
      Johannes Kepler University Linz, Research Institute for Symbolic Computation\\
      Altenberger Stra{\ss}e 69, A-4040 Linz, Austria.\\
      E-mail: \email{carsten.schneider@risc.jku.at}
}

\abstract{The large moment method can be used to compute a large number of moments of physical quantities that are described by coupled systems of linear differential equations. Besides these systems the algorithm requires a certain number of initial values as input, that are often hard to derive in a preprocessing step.
Thus a major challenge is to keep the number of initial values as small as possible.  We present the basic ideas of the underlying large moment method and present refined versions that reduce significantly the number of required initial values.}

\FullConference{14th International Symposium on Radiative Corrections (RADCOR2019)\\ 
9-13 September 2019\\
		Palais des Papes, Avignon, France}

\begin{document}

\section{Introduction}\label{Sec:Intro}

In order to solve open problems at the forefront of elementary particle physics, millions of 
complicated Feynman 
integrals in terms of a continuous parameter $x$ and the dimensional parameter $\ep$ have to be tackled. 
In order to crunch these millions of Feynman integrals to a few hundred (or thousand) so-called master integrals,
integration-by-parts (IBP) methods~\cite{Chetyrkin:1981qh,Laporta:2001dd,Studerus:2009ye,vonManteuffel:2012np,MARSEID} are applied as a preprocessing step. 
This yields an expression $\bar{f}(x,\ep)$ given as a linear combination of these master integrals. Then the main task is to compute the first coefficients of its $\ep$-expansion
\begin{equation}\label{Equ:fBarExpand}
\bar{f}(x,\ep)=\bar{f}_l(x)\ep^l+\bar{f}_{l+1}(x)\ep^{l+1}+\bar{f}_{l+2}(x)\ep^{l+2}+\dots
\end{equation}
for three-loop Feynman integrals the expansion starts usually at $l=-3$. 
More precisely, one computes for each master integral in $\bar{f}(x,\ep)$ such an $\ep$-expansion and assembles afterwards the sub-results accordingly to get the coefficients in~\eqref{Equ:fBarExpand}.
As observed in~\cite{Kotikov:1990kg,Bern:1992em,Remiddi:1997ny,Gehrmann:1999as} most of these master integrals, say $f_i(x,\ep)$, can be determined as solutions of coupled systems of linear differential equations which are of the form
\begin{eqnarray}
\label{eq:DEQEp}
\hspace*{2cm}\Dx 
\left(
\begin{smallmatrix}
f_1(x,\ep)\\ \vdots \\ f_{\lambda}(x,\ep)\end{smallmatrix}\right) 
= A
\left(\begin{smallmatrix} f_1(x,\ep)\\ \vdots \\ f_{\lambda}(x,\ep)\end{smallmatrix}\right) 
+ \left(\begin{smallmatrix} g_1(x,\ep)\\ \vdots \\ g_{\lambda}(x,\ep)\end{smallmatrix}\right),
\end{eqnarray}
with $A(x,\ep)$ being an invertible $\lambda\times\lambda$ matrix with entries from the polynomial 
ring
$\KK[x,\ep]$ in the variables $x$ and $\ep$ with coefficients from a field\footnote{We suppose that $\KK$ is computable and contains the rational numbers $\QQ$ as a sub--field.} $\KK$. Here the components $g_i(x,\ep)$ are given as a $\KK(x,\ep)$-linear combination of simpler master integrals whose $\ep$-expansions can be computed (either by solving again a coupled system or by using, e.g., symbolic summation/integration tools~\cite{Schneider:07a,Schneider:13a,Schneider:16b}). 

In the last years we have developed general algorithms in~\cite{Bluemlein:2014qka,CoupledSys:15,Coupled:16,Schneider:16b,FormFactorTwoLoops} that enable one to solve such systems in terms of indefinite nested sums (and integrals) provided that the inhomogeneous components can be represented in this class. Successful applications of these algorithms to non-trivial physical quantities can be found, e.g., in~\cite{Physics1,Physics2,FormFactorTwoLoops,PhysicsNew3}. For certain instances one may also apply methods introduced in~\cite{Henn:2013pwa,Lee:2014ioa}.

In the last years we entered more and more complicated physical problems where the above technologies are too expensive or are not anymore applicable. More precisely, solving these underlying systems recursively leads to many complicated indefinite nested sums (resp.\ indefinite nested integrals) and the calculation time explodes. Even worse, for more complicated physical problems the arising master integrals cannot be expressed anymore in terms of indefinite nested sums (or integrals) but in terms of more complicated sums/integrals, where the simpler cases come, e.g., from the class of elliptic or modular functions/forms, $_2F_1$ solutions~\cite{Broadhurst:1993mw,Bloch:2013tra,Adams:2015ydq,Remiddi:2016gno,Adams:2016xah,Ablinger:2017ptf} and general iterative-non-iterative integrals~\cite{Elliptic:18}.
However, combining all these sub-results to the final expression $\bar{f}(x)$ most of these sums/integrals vanish and only few sums/functions remain. In particular for the coefficients $\bar{f}_l(x,\ep),\dots,\bar{f}_{-1}(x,\ep)$ in the $\ep$-expansion~\eqref{Equ:fBarExpand} one often expects only indefinite nested sums/integrals, like harmonic sums~\cite{Bluemlein:99,Vermaseren:99} / harmonic polylogarithms~\cite{Remiddi:1999ew}. In other cases, only very few of these more complicated special functions remain and one wants to filter out all those sub-expression (coming from various color-factors) that are still expressible within the class of indefinite nested sums. 

Exactly for such types of problems the recently proposed large moment method~\cite{Blumlein:2017dxp} implemented in \texttt{SolveCoupledSystem}~\cite{Bluemlein:2014qka,CoupledSys:15,Coupled:16,Schneider:16b} can support the user. It allows to compute for a very large $\mu$, say $\mu=8000$, the coefficients $\bar{F}_k(0),\dots,\bar{F}_k(\mu)$ with $l\leq k\leq r$ of its power series
\begin{equation}\label{Equ:PhysicalMoment}
\bar{f}_k(x)=\sum_{n=0}^{\infty}\bar{F}_k(n)x^n.
\end{equation}
We remark that standard procedures like {\tt Mincer} \cite{Gorishnii:1989gt,Mincer:91} or {\tt MATAD} \cite{Steinhauser:2000ry} allow the calculation of a comparable small number of Mellin 
moments, e.g., $\mu=20$. 
Given such a large amount of moments, one can utilize guessing packages like~\cite{GSAGE} in order to compute recurrences for the physical quantity $\bar{F}_k(n)$.
Note that the found recurrences describe precisely the desired solution $\bar{F}_k(n)$ (i.e., the recurrence order is minimal).
Finally, one can apply our recurrence solving tools from Section~\ref{Sec:LinearRec} to hunt for representations in terms of indefinite nested sums. For instance, as demonstrated in~\cite{BKKS:09}, we could calculate from about $\mu=5000$ moments all the recurrences that determine the massless
unpolarized 3-loop anomalous dimensions and Wilson coefficients in deep-inelastic scattering
\cite{Moch:04,Vogt:2004mw,Vermaseren:2005qc}. Furthermore, we could solve all the obtained recurrences and expressed the physical quantities in terms of harmonic sums.
Similarly, we could recalculate the three-loop splitting functions~\cite{CoupledSys5} which contribute to massive operator matrix elements abinitio.

In this article we will present in details refinements of the large moment method~\cite{PhysicsNew2}. They enabled us to reproduce the polarized three-loop anomalous dimensions in~\cite{PhysicsNew1} by producing up to $\mu=6000$ moments, guessing all recurrences of the moments and solving them all. Furthermore, using our refined large moment method we could produce big parts of the heavy fermion contributions of the massive three loop form factors in~\cite{PhysicsNew2}. Here we produced up to $\mu=8000$ moments, guessed all recurrences and solved most of the recurrences yielding the desired sum representations. More precisely, all contributions up to the order $\ep^{-1}$ and big parts of the constant terms could be expressed in terms of indefinite nested sums. This also applies to the first order contributions for the massive three-loop operator matrix element $A_{Qg}^{(3)}$, see~\cite{Ablinger:2017ptf}.

The article is organized as follows.
First, we will recall our recurrence solving machinery in Section~\ref{Sec:LinearRec}. Using this technology we will elaborate in Section~\ref{Sec:SolvingCoupledSys} how one can solve coupled systems of the form~\eqref{eq:DEQEp} in terms indefinite nested sums. Finally we will deviate these ideas in Section~\ref{Sec:LM} to extract our large moment machinery~\cite{Blumlein:2017dxp}. In particular, we will present key ideas from~\cite{PhysicsNew2} to obtain refined versions that allowed us to perform the recent calculations in~\cite{PhysicsNew2,PhysicsNew1}. A conclusion is given in Section~\ref{Sec:Conclusion}.

\section{Solving linear recurrence relations in terms of indefinite nested sums}\label{Sec:LinearRec}

In the following we recall the basic ideas to solve linear recurrence relations within the class of indefinite nested sums defined over hypergeometric products\footnote{
Note that the class of indefinite nested sums contains as special cases harmonic sums~\cite{Bluemlein:99,Vermaseren:99}, cyclotomic harmonic sums~\cite{ABS:11}, generalized harmonic sums~\cite{Moch:02,ABS:13}, and nested binomials sums~\cite{ABRS:14}.}.

\begin{definition}
A product $\prod_{j=l}^kf(j)$, $l\in\NN$, is called {\em hypergeometric in $k$ over $\KK$} if $f(j)$ is an element from the rational function field $\KK(j)$ where the numerator and denominator of
$f(j)$ are nonzero if one replaces $j$ by an integer $\lambda\in\NN$ with $\lambda\geq l$.
An {\em expression in terms of indefinite nested sums over hypergeometric products in $k$ over $\KK$} is composed recursively by the three operations 
($+,-,\cdot$) with

\vspace*{-0.3cm}

\begin{itemize}
\item elements from the rational function field $\KK(k)$,

\vspace*{-0.2cm}

\item hypergeometric products in $k$ over $\KK$,

\vspace*{-0.2cm}

\item and sums of the form $\sum_{j=l}^kf(j)$ with $l\in\NN$ where 
$f(j)$ is an expression in terms of indefinite nested sums over hypergeometric products in $j$ over $\KK$; here it is assumed that the evaluation of $f(j)|_{j\mapsto\lambda}$ for all $\lambda\in\ZZ$ with $\lambda\geq l$ does not introduce any poles.
\end{itemize}

\vspace*{-0.2cm}

\noindent If~$\KK$ and $k$ are clear from the context, we call an expression in terms of indefinite nested sums defined over hypergeometric products in $k$ over $\KK$ also an {\em expression in terms of indefinite nested sums}.
\end{definition}

\noindent\textbf{Basic recurrence solving.} Suppose that we are given polynomials $a_0(x),\dots,d_d(x)\in\KK[x]$ where $a_d(n)\neq0$ for all $n\geq\delta$ for some $\delta\in\NN$ and we are given an expression $b(n)$ in terms of indefinite nested sums. Consider the sequence $(F(n))_{n\geq0}\in\KK^{\NN}$ that is defined by the initial values $F(i)=c_i\in\KK$ for $0\leq i\leq\max(\delta,d)-1$ and the linear recurrence
\begin{equation}\label{Equ:RecNoEp}
 a_0(n)F(n)+a_1(n)F(n+1)+\dots+a_d(n)F(n+d)=b(n).
\end{equation}
Then using the summation package~\texttt{Sigma}~\cite{Schneider:07a,Schneider:13a,Schneider:13b} one can \textit{decide} constructively if $F(n)$ can be calculated, up to finitely many start values, by an expression in terms of indefinite nested sums.

\medskip

Internally, this problem can be rephrased in the setting of difference rings~\cite{DR1,DR2,DR3,DR4}, and the problem can be decided afterwards in this setting using the algorithms from~\cite{Abramov:89a,Petkov:92,vanHoeij:99,Singer:99,Schneider:05a,Schneider:08c,Schneider:15}. More precisely, if these algorithms fail to find a solution, one obtains a proof that the sequence $F(n)$ cannot be represented within the class of indefinite nested sums. Otherwise, the solution in the setting of difference rings can be translated back yielding an explicit solution in terms of indefinite nested sums.

\begin{remark}\label{Remark:LMEpFree}
Suppose that $b(n)$ is not given as an expression in terms of indefinite nested sums, but only by a large number of moments, say, $b(0),b(1),\dots,b(\mu)$. Then the recurrence~\eqref{Equ:RecNoEp} together with the initial values $F(i)=c_i\in\KK$ with $0\leq i\leq\max(\delta,d)-1$ enable one to compute in linear time the moments $F(0),F(1),\dots, F(\mu)$.
\end{remark}

\noindent\textbf{$\ep$-recurrence solving.} More generally, suppose that one is given a recurrence of the form
\begin{equation}\label{Equ:RecEp}
 a_0(n,\ep)F(n,\ep)+a_1(n,\ep)F(n+1,\ep)+\dots+a_d(n,\ep)F(n+d,\ep)=b(n,\ep)
\end{equation}
with multivariate polynomials $a_i(x,\ep)\in\KK[x,\ep]$ for $0\leq i\leq d$ (not all zero) and where the inhomogeneous part can be given by a formal Laurent series expansion
\begin{equation}\label{Equ:InHomEp}
b(n,\ep)=b_l(n)\ep^{l}+b_{l+1}(n)\ep^{l+1}+b_{l+2}(n)\ep^{l+2}+\dots
\end{equation}
for some $l\in\ZZ$ where at least the coefficients $b_l(n),\dots,b_{r}(n)\in\KK$ can be computed for $n\in\NN$ by expressions in terms of indefinite nested sums.
We may suppose that not all $a_0(x,0),\dots,a_d(x,0)$ are zero. Otherwise we find a common factor $\ep^u$ with $u\geq1$ and can divide~\eqref{Equ:RecEp} by $\ep^u$. Let $d'\in\NN$ be maximal such that $a_{d'}(x,0)\neq0$ and take the smallest $\delta\in\NN$ such that   
$a_{d'}(n,0)\neq0$ for all $n\geq\delta$.\\
Then following the construction from~\cite{BKSF:12} implemented in~\texttt{Sigma} one can \textit{decide} constructively if there is a solution 
\begin{equation}\label{Equ:GeneralEpSol}
F(n,\ep)=F_l(n)\ep^{l}+F_{l+1}(n)\ep^{l+1}+F_{l+2}(n)\ep^{l+2}+\dots
\end{equation}
of~\eqref{Equ:RecEp} with the given initial values $F_j(i)=c_{j,i}\in\KK$ for $l\leq j\leq r$ and $0\leq i\leq\max(d',\delta)-1$ such that $F_j(n)$ can be calculated by an expression in terms of indefinite nested sums.

Internally, 
one proceeds as follows. Plugging in~\eqref{Equ:GeneralEpSol} into~\eqref{Equ:RecEp} and comparing coefficients at $\ep^l$ yield the constraint
\begin{equation}\label{Equ:Ep0Constraint}
a_0(n,0)F_l(n)+a_1(n,0)F_l(n+1)+\dots+a_{d'}(n,0)F_l(n+d')=b_l(n).
\end{equation}
Together with the initial values $F_l(i)=c_{l,i}$ for $0\leq i\leq\max(d',\delta)-1$ one can utilize \texttt{Sigma} (see above) to  decide algorithmically if $F_l(n)$ can be represented in terms of indefinite nested sums. If this is not possible, the algorithm stops. Otherwise, one obtains such an expression for $F_l(n)$ and one gets
\begin{equation}\label{Equ:recNext}
a_0(n,\ep)F'(n,\ep)+a_1(n,\ep)F'(n+1,\ep)+\dots+a_{d}(n,\ep)F'(n+d,\ep)=b'(n,\ep)
\end{equation}
with the updated right hand side 
\begin{equation}\label{Equ:FormulaForbPrime}
b'(n,\ep)=b(n,\ep)-\Big(a_0(n,\ep)F_l(n)+a_1(n,\ep)F_l(n+1)+\dots+a_d(n,\ep)F_l(n+d)\Big)
\end{equation}
and the unknown $\ep$-expansion
\begin{equation}\label{Equ:FPrime}
F'(n,\ep)=F_{l+1}(n)\ep^{l+1}+F_{l+2}(n)\ep^{l+2}+\dots
\end{equation}
Note that the first $r-l$ coefficients in
\begin{equation}\label{Equ:bPrime}
b'(n,\ep)=b'_{l+1}(n)\ep^{l+1}+b'_{l+2}(n)\ep^{l+2}+\dots,
\end{equation}
i.e., in the $\ep$-expansion of~\eqref{Equ:FormulaForbPrime} 
can be given in terms of indefinite nested sums. Thus we can repeat the above strategy and can decide algorithmically if the remaining coefficients $F_{l+1}(n),\dots,F_{r}(n)$ can be represented in terms of indefinite nested sums.

\begin{remark}
Note that the above algorithm computes the maximal $\lambda\in\NN$ with $l-1\leq\lambda\leq r$ such that $F_{l}(n),\dots,F_{r}(n)$ can be represented in terms of indefinite nested sums and returns the coefficients $F_{l}(n),\dots,F_{\lambda}(n)$ in such a representation.  
\end{remark}

\begin{remark}\label{Remark:LMEp}
Suppose that the coefficients $b_i(n)$ in~\eqref{Equ:InHomEp} are not given as expressions in terms of indefinite nested sums, but only by a large number of moments, say, $b_j(0),b_j(1),\dots,b_j(\mu)$. Then the recurrence~\eqref{Equ:RecNoEp} together with the initial values $F_j(i)=c_{j,i}\in\KK$ with $0\leq i\leq\max(\delta,d')-1$ enables one to compute in linear time the moments $F_j(0),F_j(1),\dots, F_j(\mu)$. More precisely, one starts with $j=l$, obtains the recurrence~\eqref{Equ:Ep0Constraint} with the moment $b_l(0),b_l(1),\dots,b_l(\mu)$ and computes the moments $F_l(0),F_l(1),\dots,F_l(\mu)$; see Remark~\ref{Remark:LMEpFree}. Then one computes the moments for $b'_{j}(n)$ in~\eqref{Equ:bPrime} using the formula~\eqref{Equ:FormulaForbPrime}.  This yields~\eqref{Equ:recNext} with~\eqref{Equ:FPrime} and we are in the position to repeat this process. Namely, we can calculate iteratively the moments $F_{j}(0),F_{j}(1),\dots,F_{j}(\mu)$ for $j=l+1,l+2,\dots,r$.
\end{remark}

\section{Solving coupled systems of first-order linear differential equations}\label{Sec:SolvingCoupledSys}

In the following we present our main tools to solve coupled systems in terms of indefinite nested sums.

\medskip

\noindent\textbf{Coupled system solving.} Suppose that we are given a coupled system of the form~\eqref{eq:DEQEp}
with $A(x,\ep)$ being an invertible $\lambda\times\lambda$ matrix with entries from the polynomial 
ring $\KK[x,\ep]$. Furthermore, suppose that the inhomogeneous parts can be given in form of a power series 
\begin{equation}\label{Equ:GiDef1}
g_i(x,\ep)=\sum_{n=0}^{\infty}G_i(n,\ep)x^n
\end{equation}
where the coefficients itself can be given in form of the $\ep$-expansions
\begin{equation}\label{Equ:GiDef2}
G_i(n,\ep)=G_{i,l}(n)\,\ep^{l}+G_{i,l+1}(n)\,\ep^{l+1}+G_{i,l+2}(n)\,\ep^{l+2}\dots
\end{equation}
If the coefficients $G_{i,k}(n)$, free of $x$ and $\ep$, for $l\leq k\leq r'_i$ ($r'_i$ sufficiently high) can be represented in terms of indefinite nested sums, one can decide algorithmically if also the unknown functions $f_1(x,\ep),\dots, f_{\lambda}(x,\ep)$ can be given in such a form where the highest $\ep$ orders are $r_1,\dots,r_{\lambda}$, respectively. 

Here one proceeds as follows.
\begin{enumerate}
 \item By uncoupling algorithms, like, e.g., Z\"urcher's algorithm~\cite{Zuercher:94} implemented in the package \texttt{OreSys}~\cite{ORESYS}, one obtains a scalar linear differential equation of the form
 \begin{equation}\label{Equ:UncoupledEq}
 \alpha_0(x,\ep)f_1(x,\ep)+\alpha_1(x,\ep)D_x\,f_1(x,\ep)+\dots+ \alpha_{\lambda}(x,\ep)D_x^{\lambda}f_1(x,\ep)=\beta(x,\ep)
 \end{equation}
 with
 $$\beta(x,\ep)=\sum_{i,j\geq0}\beta_{i,j}(x,\ep)D_x^ig_j(x,\ep)$$ 
 for explicitly given $\beta_{i,j}\in\KK(x,\ep)$; note that only finitely many $\beta_{i,j}$ are non-zero.
 In addition, one gets
 \begin{equation}\label{Equ:fKFormula}
 f_k(x,\ep)=\sum_{k,i\geq0}\phi_{k,i}(x,\ep)\Dx^if_1(x,\ep),\quad 2\leq k\leq\lambda,
\end{equation}
 for explicitly given $\phi_{k,i}\in\KK(x,\ep)$; note that only finitely many $\phi_{k,i}$ are non-zero
\item Plugging in the ansatz 
\begin{equation}\label{Equ:F1XExp}
f_1(x,\ep)=\sum_{n=0}^{\infty}F_{1}(n,\ep)x^n
\end{equation}
into~\eqref{Equ:UncoupledEq} and performing coefficient comparison w.r.t.\ $x^n$ yield a recurrence of the form~\eqref{Equ:RecEp} (with $F(n,\ep)$ replaced by $F_1(n,\ep)$) for some explicitly given $a_0(x,\ep),\dots,a_d(x,\ep)\in\KK[x,\ep]$ and~\eqref{Equ:InHomEp} where the first coefficients $b_i(n,\ep)$ can be represented in terms of indefinite nested sums; note that the orders $r'_i$ of the $\ep$-expansions in~\eqref{Equ:GiDef2} must be set high enough to obtain the correct expressions for $b_i(n,\ep)$ up to the order $r_1$. 

\noindent Note further that the recurrence order $d\in\NN$ of~\eqref{Equ:RecEp} is bounded by
\begin{equation}\label{Equ:OrderBound}
d\leq \lambda+\max_{0\leq i\leq \lambda}\deg_x(\alpha_i).
\end{equation}
\item Applying the tools from Section~\ref{Sec:LinearRec} together with the corresponding initial values one can decide if the first coefficients $F_{1,i}(n)$ of the $\ep$-expansion
\begin{equation}\label{Equ:F1Exp}
F_{1}(n,\ep)=F_{1,l}(n)\ep^l+F_{1,l+1}(n)\ep^{l+1}+F_{1,l+2}(n)\ep^{l+2}+\dots
\end{equation}
can be given in terms of indefinite nested sums. If this fails, stop.
\item Plugging the $\ep$-expansion~\eqref{Equ:F1Exp} (where the first coefficients are given explicitly) into~\eqref{Equ:fKFormula} and extracting the coefficient of $x^n$ yield
\begin{equation}\label{Equ:FkRep}
F_{k}(n,\ep)=F_{k,l}(n)\ep^l+F_{k,l+1}(n)\ep^{l+1}+F_{k,l+2}(n)\ep^{l+2}+\dots,\quad 2\leq k\leq\lambda
\end{equation}
where the first coefficients $F_{k,i}(n)$ can be represented in terms of indefinite nested sums; note that the orders $r_1,r'_1,\dots,r'_{\lambda}$ of the $\ep$-expansions $F_1(n,\ep),G_1(n,\ep),\dots,G_{\lambda}(n,\ep)$ must be set sufficiently high to get the desired $\ep$-expansions of $F_k(n,\ep)$ up to the orders $r_k$.
\end{enumerate}

\noindent We note that this machinery has been implemented efficiently in the package \texttt{SolveCoupled\-System}~\cite{Bluemlein:2014qka,CoupledSys:15,Coupled:16,Schneider:16b} that relies on the following sub-packages:

\vspace*{-0.2cm}

\begin{enumerate}
 \item \texttt{Sigma}~\cite{Schneider:07a,Schneider:13a,Schneider:13b} to find the solutions of the recurrence~\eqref{Equ:RecEp}.

 \vspace*{-0.2cm}

 \item \texttt{HarmonicSums}~\cite{Bluemlein:99,Vermaseren:99,Ablinger:10,Ablinger:12,ABS:11,ABS:13,ABRS:14,Ablinger:2014dda,Ablinger:2017Mellin} to support \texttt{Sigma} for the elimination of algebraic relations among the arising sums using quasi-shuffle algebras~\cite{Hoffman,Bluemlein:04,AS:18}.

 \vspace*{-0.2cm}

 \item \texttt{SumProduction}~\cite{Schneider:13b} to obtain from~\eqref{Equ:fKFormula} the coefficients in~\eqref{Equ:FkRep} in terms of indefinite nested sums.
\end{enumerate}

\vspace*{-0.2cm}

\noindent Using all these technologies we succeeded in calculating, e.g., various non-trivial physical quantities in~\cite{CoupledSys1,CoupledSys2}. 
We remark further, that one can apply also differential equation solvers (available e.g.\ in \texttt{HarmonicSums}, see~\cite{Ablinger:2017Mellin}) to~\eqref{Equ:UncoupledEq} in order to find closed form solutions in terms of indefinite nested integrals defined over hyperexponential functions. For more details we refer to~\cite{FormFactorTwoLoops,PhysicsNew3}.

\section{The large moment method and refined variants}\label{Sec:LM}

Suppose that we utilized IBP methods~\cite{Chetyrkin:1981qh,Laporta:2001dd,Studerus:2009ye,vonManteuffel:2012np,MARSEID} and obtained a physical expression $\bar{f}(x,\ep)$ in terms of master integrals that are solutions of coupled systems of linear differential equations of the form~\eqref{eq:DEQEp}. A natural tactic is then to solve these coupled systems, e.g., in terms of indefinite nested sums as described in Section~\ref{Sec:SolvingCoupledSys} and to combine the solutions to obtain the first coefficients of the $\ep$-expansion~\eqref{Equ:fBarExpand}; here the  coefficients $\bar{f}_k(x)$ itself are considered in power series expansions~\eqref{Equ:PhysicalMoment} and one seeks for an all-$n$ solution of the corresponding coefficients $\bar{F}_k(n)$.

As mentioned in the introduction, master integrals may pop up that cannot be expressed in terms of indefinite nested sums. In particular, the inhomogeneous components in~\eqref{eq:DEQEp} and thus the coefficients in~\eqref{Equ:InHomEp} of the underlying recurrence~\eqref{Equ:RecEp} cannot be expressed in terms of indefinite nested sums, or the recurrence cannot be solved within this class. 
In order to overcome this situation, we have introduced the large moment machinery in~\cite{Blumlein:2017dxp} that enables one to compute the moments $\bar{F}_k(n)$ with $n=0,1,\dots,\mu$ for a large number $\mu\in\NN$ of the power series expansion~\eqref{Equ:PhysicalMoment}. This enables one to analyze this physical quantity further by numerical methods. In addition, one can try to compute linear recurrences by guessing methods~\cite{GSAGE} and to solve them as described in Section ~\ref{Sec:LinearRec}; for concrete calculations see also the end of Section~\ref{Sec:Intro}.

\medskip

Internally, the large moment machinery works as follows.

\smallskip

\noindent\textbf{The original large moment method.}
Suppose that we are given a coupled system of the form~\eqref{eq:DEQEp} where the inhomogeneous part $g_i(x,\ep)$ has the coefficients $G_{i,k}(n)$ in~\eqref{Equ:GiDef1}. But the coefficients are not given as an all-$n$ representation (e.g., in terms of indefinite nested sums as assumed in Section~\ref{Sec:SolvingCoupledSys}), but by a finite number of moments $G_{i,k}(0),G_{i,k}(2),\dots,G_{i,k}(\mu)$ where $\mu$ is large (e.g., $\mu=8000$). They can be determined, e.g., by 

\vspace*{-0.3cm}

\begin{itemize}
\item other coupled systems to which the large moment method is applied recursively;

\vspace*{-0.2cm}

\item symbolic summation or integration methods~\cite{Schneider:07a,Schneider:13a,Schneider:16b} that yield 
representations in terms of indefinite nested sums or integrals from which one can produce a large number of moments;

\vspace*{-0.2cm}

\item by standard procedures like {\tt Mincer} \cite{Gorishnii:1989gt,Mincer:91} or {\tt MATAD} \cite{Steinhauser:2000ry} 
if only a small 
number of moments contributes. Also analytically solvable (multiple) Mellin-Barnes representations can be used in some cases.
\end{itemize}

\vspace*{-0.2cm}

Given this input, one follows the calculation steps given in Section~\ref{Sec:SolvingCoupledSys}. But instead of dealing with expressions in terms of indefinite nested sums, one performs the calculation steps for lists with entries from $\KK$ that encode the moments of the underlying expressions. Adapting this procedure, one obtains a recurrence~\eqref{Equ:RecEp} (with $F(n,\ep)$ replaced by $F_1(n,\ep)$ given in~\eqref{Equ:F1Exp}) 
where the polynomials $a_i(x,\ep)\in\KK[x,\ep]$ are given explicitly but where coefficients $b_i(n)$ in~\eqref{Equ:InHomEp} are not represented as expressions in terms of indefinite nested sums, but are given explicitly by the moments $b_i(0),b_i(1),\dots,b_i(\mu)$. Thus using the first initial values we can calculate the moments $F_{1,k}(0),F_{1,k}(1),\dots,F_{1,k}(\mu)$ for $k=l,l+1,\dots,r_1$ in~\eqref{Equ:F1Exp} by Remark~\ref{Remark:LMEp}.\\
Finally, we can calculate the moments $F_{j,k}(0),F_{j,k}(1),\dots,F_{j,k}(\mu)$ of~\eqref{Equ:FkRep} for $j=2,\dots,\lambda$ and $k=l,\dots,r_j$ using the formulas~\eqref{Equ:fKFormula} and the moments $F_{1,k}(0),\dots,F_{1,k}(\mu)$ for $k=l,\dots,r_1$.

\medskip

This general machinery introduced in~\cite{Blumlein:2017dxp} worked successfully, e.g., for the calculation of the splitting functions~\cite{CoupledSys5}. But to carry out larger problems such as~\cite{PhysicsNew1,PhysicsNew2} we faced the following 

\medskip

\noindent\textbf{Bottleneck.} In order to execute the large moment machinery, sufficiently many initial values must be provided in a preprocessing step. In general, the number of initial values equals the order $d$ of the recurrence~\eqref{Equ:recNext}
(sometimes it might be reduced if $a_d(x,0)=0$ and it might be increased if $a_d(x,0)\neq0$ but $a_d(x,d)=0$). In particular, the number of initial values is bounded by~\eqref{Equ:OrderBound} (and by experiences it often comes close to this bound). For simpler situations like in~\cite{CoupledSys5} the above approach was completely sufficient. However, e.g., for systems coming from~\cite{PhysicsNew1,PhysicsNew2} the coefficients $\alpha_0(x,\ep),\dots,\alpha_{\lambda}(x,\ep)$ have rather high degrees (e.g., up to 50) and thus also the recurrence order is of similar magnitude.
Unfortunately, the calculation of this amount of initial values is often too hard and thus the above large moment machinery is out of scope.

\medskip

Based on these observations we succeeded in~\cite{PhysicsNew2} in reducing the required initial values significantly to a number that could be computed in reasonable time (or that have been calculated already in earlier projects). We remark that this improvement has its price:  the calculation of the moments gets more involved. Nevertheless, a slower method is better than a method that cannot be applied due to missing initial values.

\medskip

\noindent\textbf{Refinement 1.} Suppose that the greatest common divisor of the coefficients $\alpha_{0}(x,\ep),\dots,\alpha_{\lambda}(x,\ep)$ in~\eqref{Equ:UncoupledEq} is $1$, i.e., there is no common polynomial factor that depends on $x$ or $\ep$. 
This implies that there is at least one $i$ such that $\alpha_{i}(x,0)\neq0$. 
By coefficient comparisons w.r.t.\ $\ep^l$ and $x^n$ (and appropriate expansions) one gets directly a recurrence of the form~\eqref{Equ:RecNoEp} (with $F(n)$ replaced by $F_{1,l}(n)$). Analogously to the method in Section~\ref{Sec:LinearRec} one can now repeat the game: Given sufficiently many initial values, one can calculate the moments $F_{1,l}(0),F_{1,l}(1),\dots,F_{1,l}(\mu)$. Afterwards, one plugs~\eqref{Equ:F1XExp} with~\eqref{Equ:F1Exp}
into~\eqref{Equ:UncoupledEq}, updates the inhomogeneous side $\beta(x,\ep)$ and obtains the differential equation~\eqref{Equ:UncoupledEq} (with modified $\beta(x,\ep)$) where $f_1(x,\ep)$ takes over the role of 
$$f'_1(x,\ep)=f_{1,l+1}(x)\ep^{l+1}+f_{1,l+2}(x)\ep^{l+2}+\dots.$$ 
Thus we can loop up to get $F_{1,k}(0),F_{1,k}(1),\dots,F_{1,k}(\mu)$ with $k=l+1,\dots,r_1$. Afterwards, one computes the remaining coefficients $F_{j,k}(0),\dots,F_{j,k}(\mu)$ with $j=2,\dots,\lambda$ and $1\leq k\leq r_j$ as described above. 

\medskip

\noindent A benefit of this modified procedure is that the uncoupling method does not have to deal with $\ep$, i.e., the rational function arithmetic is reduced from $\KK(x,\ep)$ to $\KK(x)$. We note that the obtained recurrence contains only the information for one coefficient $F_{1,k}(n)$, while the recurrence of the original method (depending on $\ep$) contains the information to describe the full $\ep$-expansion with the coefficients $F_{1,l}(n),F_{1,l+1}(n),\dots$ -- this extra information often requires a recurrence of higher order. Another major improvement is obtained by refining this approach further with 

\medskip

\noindent\textbf{Refinement 2.} 
Let $p(x)\in\KK[x]\setminus\{0\}$ be the 
greatest common divisor of $\alpha_{0}(x,0),\dots,\alpha_{\lambda}(x,0)$, i.e., $p(x)$ contains all common polynomial factors in 
$x$ of the $\alpha_{i}(x,0)$. 
Dividing~\eqref{Equ:UncoupledEq} by $p(x)$ yields
\begin{equation}\label{Equ:NormalizedEq}
\sum_{i=0}^{\lambda}\frac{\alpha_{i}(x,\ep)}{p(x)}\Dx^{i} f_1(x,\ep)=\frac{\beta(x,\ep)}{p(x)}.
\end{equation}
Performing the expansions w.r.t.\ $x$ and $\ep$ on the right-hand side and doing coefficient comparison w.r.t.\ $\ep^l$ and $x^j$ afterwards yield a linear recurrence of the form~\eqref{Equ:RecNoEp} (with $F(n)$ replaced by $F_{1,l}(n)$) where the order $d$ is bounded by
\begin{equation}\label{Equ:ReducedUpperBound}
d\leq \lambda+\max_{0\leq i\leq \lambda}\deg_x(\alpha_i)-\deg_x(p).
\end{equation}
After the computation of the moments $F_{1,l}(0),F_{1,l}(1),\dots,F_{1,l}(\mu)$ with that recurrence we proceed as in Refinement~1 incorporated the ideas of Refinement~2 in each step.

\medskip

\noindent As it turns out, the degree of the common factor $p(x)$ of the coefficients $\alpha_i(x,0)$ with $0\leq i\leq\lambda$ is often rather large and thus also the upper bound of the recurrence order can be reduced significantly; see~\eqref{Equ:ReducedUpperBound}. Even better, the recurrence order $d$ is reduced by the value $\deg_x(p)$ and therefore the number of required initial values can be reduced substantially; see Example~\ref{Exp:Example} below.

\medskip

Using, e.g., Z\"urcher's algorithm~\cite{Zuercher:94}, one obtains a fully decoupled system with usually one scalar linear differential equation~\eqref{Equ:UncoupledEq} for the unknown $f_1(x,\ep)$ and further formulas~\eqref{Equ:fKFormula} that enable one to express the remaining $f_k(x,\ep)$ in terms of $f_1(x,\ep)$. Alternatively, one may take the simple Gauss method (available in~\texttt{OreSys}~\cite{ORESYS}) that yields the following 

\medskip

\noindent\textbf{Refinement 3.} One obtains a linear differential equation of $f_{\lambda}(x,\ep)$ of order $o_{\lambda}$ where the inhomogeneous part depends only on the inhomogeneous components $g_i(x,\ep)$ of~\eqref{eq:DEQEp} together with the application of $D_x^u$ with $u\in\NN$ on these components. Next, one obtains a linear differential equation for $f_{\lambda-1}(x,\ep)$ of order $o_{\lambda-1}$ whose inhomogeneous part depends additionally on $D_x^uf_{\lambda-1}(x,\ep)$ with $u\in\NN$. More generally, one gets a linear differential equation for $f_{j}(x,\ep)$ of order $o_j$ with $1\leq j\leq\lambda$ whose inhomogeneous part depends on the components $g_1(x,\ep),\dots,g_{\lambda}(x,\ep)$ and additionally on the unknowns $f_{j+1}(x,\ep),\dots,f_{\lambda}(x,\ep)$ and the application of $D_x^u$ with $u\in\NN$ to them. Thus one can compute iteratively the unknowns $f_j(x,\ep)$ for $j=\lambda,\lambda-1,\dots,1$ applying Refinement~2. This extra work is rewarded by the fact that each of the orders $o_1,\dots,o_{\lambda}$ is usually much smaller than $\lambda$ which implies that for each $F_{j,k}(n)$ the number of initial values is reduced. In particular, we get the improved upper bound
$$d\leq o_i+\max_{0\leq i\leq \lambda}\deg_x(\alpha_i)-\deg_x(p).$$ 
Note that the total number of initial values needed for all components $F_{j,k}(n)$ might be even larger. However, it is usually much harder to calculate initial values $F_{1,k}(i)$ for large values $i$ (as proposed in the earlier strategies) than computing initial values for more components $F_{j,k}(i)$ with $1\leq j\leq\lambda$ where $i$ is small.

\medskip

\begin{example}\label{Exp:Example}
 Consider a typical system~\eqref{eq:DEQEp} with $\lambda=4$ coming from~\cite{PhysicsNew2}.
For the uncoupled system we obtain 4 linear differential equations of orders $o_1=4$, $o_2=2$, $o_3=0$ and $o_4=0$.
Here the inhomogeneous part of the linear differential equation in $f_k(x,\ep)$ of order $o_k$ depends on the components $D_x^ug_1(x,\ep)$, $D_x^ug_2(x,\ep)$, $D_x^ug_3(x,\ep)$, $D_x^ug_4(x,\ep)$ and $D_x^uf_{k+1}(x,\ep),\dots,D_x^uf_{4}(x,\ep)$ with $u\in\NN$. Thus we compute the moments stepwise for $f_{4}(n,\ep)$, $f_{3}(n,\ep)$, $f_{2}(n,\ep)$, and finally for $f_{1}(n,\ep)$. More precisely, we compute the moments of the coefficients $F_{j,k}(n)$ in~\eqref{Equ:F1Exp} and~\eqref{Equ:FkRep} with $1\leq j\leq 4$ and $l=-3\leq k$.
E.g.,
for $F_{1,-3}(n)$ we obtain a recurrence of the form~\eqref{Equ:RecNoEp} ($F(n)$ replaced by $F_{1,-3}(n)$)
of order $d=4$. We note that this small recurrence order was possible by Refinement~2:  we sneaked in a polynomial 
$p(x)\in\QQ[x]$ of degree $13$ within the linear differential equation~\eqref{Equ:NormalizedEq}; setting $p(x)=1$ would have delivered a recurrence of order $4+13=17$. Finally using 4 initial values enabled us to compute the moments $F_{1,-3}(0),F_{1,-3}(1),\dots,F_{1,-3}(8000)$. Similarly all the other coefficients can be calculated. 
\end{example}

\section{Conclusion}\label{Sec:Conclusion}
We presented our general large moment method introduced in~\cite{Blumlein:2017dxp}. In order to activate it, sufficiently many initial values have to be provided as a preprocessing step. However, the calculation of these starting points is often the show-stopper and a major challenge is to reduce the required number of initial values as much as possible. In particular, the maximum of the number of initial values among all components $F_{j,k}(n)$ should be kept as small as possible.  In this article we have elaborated three refinements from~\cite{PhysicsNew2} that improved this problem significantly. This new large moment engine enabled us to recalculate the polarized three-loop anomalous dimensions in~\cite{PhysicsNew1} and to tackle big parts of the heavy fermion contributions of the massive three loop form factors in~\cite{PhysicsNew2}. We remark that additional improvements have been introduced in~\cite{PhysicsNew2} that require further investigations.

\vspace{2ex}
\noindent
{\bf Acknowledgment.} This work was supported in part by the Austrian Science Fund (FWF) grant SFB F50 (F5009-N15), by the 
EU TMR network SAGEX Marie Sk\l{}odowska-Curie grant agreement No. 764850 and COST action CA16201: Unraveling new 
physics at the LHC through the precision frontier.


\begin{thebibliography}{10}

\setlength{\parskip}{0pt}
\setlength{\itemsep}{0pt plus 0ex}

\footnotesize

\bibitem{Ablinger:10}
  J.~Ablinger.
  {\em A computer algebra toolbox for harmonic sums related to particle physics}. Diploma Thesis, J. Kepler University Linz, 2009,
  arXiv:1011.1176 [math-ph].


\bibitem{Ablinger:12}
J.~Ablinger.
\newblock {\em Computer algebra algorithms for special functions in particle
  physics}.
\newblock PhD thesis, J. Kepler University Linz, April 2012.

\bibitem{Ablinger:2014dda}
  J.~Ablinger.
  The package HarmonicSums: Computer algebra and analytic aspects of nested sums,
  PoS {(LL2014)} 019;


\bibitem{Ablinger:2017Mellin}
J.~Ablinger.
\newblock Computing the inverse Mellin transform of holonomic sequences using
  {K}ovacic's algorithm.
\newblock In A.~Hoang and C.~Schneider, editors, {\em PoS (RADCOR2017) 069},
  pages 1--8, 2017.

\bibitem{Coupled:16}
J.~Ablinger, A.~Behring, J.~Bl\"umlein, A.~de~Freitas, and C.~Schneider.
\newblock {Algorithms to solve coupled systems of differential equations in
  terms of power series}.
\newblock In J.~Bl\"umlein, P.~Marquard, and T.~Riemann. (ed.), editors, {\em {Proc.
  Loops and Legs in Quantum Field Theory - LL 2016}}, volume PoS(LL2016)005,
  pages 1--15, 2016.
\newblock arXiv:1608.05376 [cs.SC].

\bibitem{CoupledSys1}
J.~Ablinger, A.~Behring, J.~Bl{\"u}mlein, A.~De~Freitas, A.~von Manteuffel, and
  C.~Schneider.
\newblock {The 3-loop pure singlet heavy flavor contributions to the structure
  function {$F_2(x,Q^2)$} and the anomalous dimension}.
\newblock {\em Nucl. Phys. B}, 890:48--151, 2014.

\bibitem{CoupledSys5}
J.~Ablinger, A.~Behring, J.~Bl{\"u}mlein, A.~De~Freitas, A.~von Manteuffel, and
  C.~Schneider.
\newblock The three-loop splitting functions {$P_{qg}^{(2)}$} and {$P_{gg}^{(2,
  N_F)}$}.
\newblock {\em Nucl. Phys. B}, 922:1--40, 2017.

\bibitem{FormFactorTwoLoops}
J.~Ablinger, A.~Behring, J.~Bl{\"u}mlein, G.~Falcioni, A.~D. Freitas,
  P.~Marquard, N.~Rana, and C.~Schneider.
\newblock The heavy quark form factors at two loops.
\newblock {\em Phys. Rev. D}, 97(094022):1--44, 2018.

\bibitem{Schneider:16b}
J.~Ablinger, A.~Behring, J.~Bl{\"u}mlein, A.~D. Freitas, A.~von Manteuffel, and
  C.~Schneider.
\newblock Calculating three loop ladder and {V}-topologies for massive operator
  matrix elements by computer algebra.
\newblock {\em Comput. Phys. Comm.}, 202:33--112, 2016.

\bibitem{Ablinger:2017ptf}
J.~Ablinger, A.~Behring, J.~Bl{\"u}mlein, A.~D. Freitas, A.~von Manteuffel, and
  C.~Schneider.
\newblock Heavy flavor {W}ilson coefficients in deep-inelastic scattering:
  Recent results.
\newblock
{\em
  PoS(QCDEV2017)031}, pages 1--10, 2017.

\bibitem{CoupledSys:15}
J.~Ablinger, J.~Bl{\"u}mlein, A.~de~Freitas, and C.~Schneider.
\newblock {A toolbox to solve coupled systems of differential and difference
  equations}.
\newblock {\em PoS (RADCOR2015) 060}, 2015,      arXiv:1601.01856 [cs.SC].

\bibitem{Physics2}
J.~Ablinger, J.~Bl{\"u}mlein, A.~D. Freitas, A.~Hasselhuhn, A.~von Manteuffel,
  M.~Round, C.~Schneider, and F.~Wi\ss{}brock.
\newblock The transition matrix element {$A_{gq}(N)$} of the variable flavor
  number scheme at {$O(\alpha_s^3)$}.
\newblock {\em Nucl. Phys. B}, 882:263--288, 2014.

\bibitem{Elliptic:18}
J.~Ablinger, J.~Bl\"umlein, A.~D. Freitas, M.~van Hoeij, E.~Imamoglu, C.~Raab,
  C.~Radu, and C.~Schneider.
\newblock Iterated elliptic and hypergeometric integrals for {F}eynman
  diagrams.
\newblock {\em J. Math. Phys.}, 59(062305):1--55, 2018.

\bibitem{PhysicsNew3}
J.~Ablinger, J.~Bl\"umlein, P.~Marquard, N.~Rana, and C.~Schneider.
\newblock Automated solution of first order factorizable systems of
  differential equations in one variable.
\newblock {\em Nucl. Phys. B}, 939:253--291, 2019.


\bibitem{ABRS:14}
J.~Ablinger, J.~Bl\"umlein, C.~Raab, and C.~Schneider.
\newblock Iterated binomial sums and their associated iterated integrals.
\newblock {\em J. Math. Phys.}, 55(112301):1--57, 2014.


\bibitem{ABS:11}
J.~Ablinger, J.~Bl\"umlein, and C.~Schneider.
\newblock Harmonic sums and polylogarithms generated by cyclotomic polynomials.
\newblock {\em J. Math. Phys.}, 52(10):1--52, 2011.

\bibitem{ABS:13}
J.~Ablinger, J.~Bl\"umlein, and C.~Schneider.
\newblock Analytic and algorithmic aspects of generalized harmonic sums and
  polylogarithms.
\newblock {\em J. Math. Phys.}, 54(8):1--74, 2013.

\bibitem{AS:18}
J.~Ablinger and C.~Schneider.
\newblock Algebraic independence of sequences generated by (cyclotomic)
  harmonic sums.
\newblock {\em Annals of Combinatorics}, 22(2):213--244, 2018.

\bibitem{Abramov:89a}
S.~Abramov.
\newblock Rational solutions of linear differential and difference equations
  with polynomial coefficients.
\newblock {\em U.S.S.R. Comput. Math. Math. Phys.}, 29(6):7--12, 1989.

\bibitem{Adams:2016xah}
L.~Adams, C.~Bogner, A.~Schweitzer, and S.~Weinzierl.
\newblock The kite integral to all orders in terms of elliptic polylogarithms.
\newblock {\em J. Math. Phys.}, 57(12):122302, 2016.

\bibitem{Adams:2015ydq}
L.~Adams, C.~Bogner, and S.~Weinzierl.
\newblock {The iterated structure of the all-order result for the two-loop
  sunrise integral}.
\newblock {\em J. Math. Phys.}, 57(3):032304, 2016.

\bibitem{PhysicsNew1}
A.~Behring, J.~Bl\"umlein, A.~D. Freitas, A.~Goedicke, S.~Klein, A.~von
  Manteuffel, C.~Schneider, and K.~Sch\"onwald.
\newblock The polarized three-loop anomalous dimensions from on-shell massive
  operator matrix elements.
\newblock {\em Nucl. Phys. B}, 948(114753):1--41, 2019.

\bibitem{CoupledSys2}
A.~Behring, J.~Bl\"umlein, A.~D. Freitas, A.~von Manteuffel, and C.~Schneider.
\newblock The 3-loop non-singlet heavy flavor contributions to the structure
  function {$g_1(x,Q^2)$} at large momentum transfer.
\newblock {\em Nucl. Phys. B}, 897:612--644, 2015.

\bibitem{Bern:1992em}
Z.~Bern, L.~Dixon, and D.~Kosower.
\newblock {Dimensionally regulated one loop integrals}.
\newblock {\em Phys. Lett.}, B302:299--308, 1993.

\bibitem{Bloch:2013tra}
S.~Bloch and P.~Vanhove.
\newblock {The elliptic dilogarithm for the sunset graph}.
\newblock {\em J. Number Theor.}, 148:328--364, 2015.

\bibitem{Bluemlein:04}
J.~Bl{\"u}mlein.
\newblock Algebraic relations between harmonic sums and associated quantities.
\newblock {\em Comput. Phys. Commun.}, 159(1):19--54, 2004.

\bibitem{Physics1}
J.~Bl\"umlein, A.~Hasselhuhn, S.~Klein, and C.~Schneider.
\newblock The {$O(\alpha_s^3 n_f T_F^2 C_{A,F})$} contributions to the gluonic
  massive operator matrix elements.
\newblock {\em Nucl. Phys. B}, 866:196--211, 2013.


\bibitem{BKKS:09}
J.~Bl{\"u}mlein, M.~Kauers, S.~Klein, and C.~Schneider.
\newblock Determining the closed forms of the {$O(a_s^3)$} anomalous dimensions
  and {W}ilson coefficients from {M}ellin moments by means of computer algebra.
\newblock {\em Comput. Phys. Commun.}, 180:2143--2165, 2009.

\bibitem{BKSF:12}
J.~Bl\"umlein, S.~Klein, C.~Schneider, and F.~Stan.
\newblock A symbolic summation approach to {F}eynman integral calculus.
\newblock {\em J. Symbolic Comput.}, 47:1267--1289, 2012.

\bibitem{Bluemlein:99}
J.~Bl\"umlein and S.~Kurth.
\newblock Harmonic sums and {M}ellin transforms up to two-loop order.
\newblock {\em Phys. Rev.}, D60: 014018, 1999.

\bibitem{PhysicsNew2}
J.~Bl\"umlein, P.~Marquard, N.~Rana, and C.~Schneider.
\newblock The heavy fermion contributions to the massive three loop form
  factors.
\newblock {\em Nucl. Phys. B}, 949(114751):1--97, 2019.


\bibitem{Blumlein:2017dxp}
J.~Bl{\"u}mlein and C.~Schneider.
\newblock The method of arbitrarily large moments to calculate single scale
  processes in quantum field theory.
\newblock {\em Phys. Lett.}, B771:31--36, 2017.

\bibitem{Broadhurst:1993mw}
D.~Broadhurst, J.~Fleischer, and O.~Tarasov.
\newblock {Two loop two point functions with masses: Asymptotic expansions and
  Taylor series, in any dimension}.
\newblock {\em Z. Phys.}, C60:287--302, 1993.

\bibitem{Chetyrkin:1981qh}
K.~Chetyrkin and F.~Tkachov.
\newblock Integration by parts: The algorithm to calculate beta functions in 4
  loops.
\newblock {\em Nucl. Phys. B}, 192:159--204, 1981.

\bibitem{Bluemlein:2014qka}
A.~De~Freitas, J.~Bl{\"u}mlein, and C.~Schneider.
\newblock Recent symbolic summation methods to solve coupled systems of
  differential and difference equations.
\newblock {\em PoS (LL2014) 017}, 2014.

\bibitem{Gehrmann:1999as}
T.~Gehrmann and E.~Remiddi.
\newblock {Differential equations for two loop four point functions}.
\newblock {\em Nucl. Phys. B}, 580:485--518, 2000.

\bibitem{ORESYS}
S.~Gerhold.
\newblock {\em Uncoupling systems of linear ore operator equations}.
\newblock Master's thesis, RISC, J. Kepler University Linz, 2 2002.

\bibitem{Gorishnii:1989gt}
  S.G.~Gorishnii, S.A.~Larin, L.R.~Surguladze and F.V.~Tkachov.
  Mincer: Program for multiloop calculations in Quantum Field Theory for the Schoonschip system,
  {\em Comput.\ Phys.\ Commun.},  55: 381--408, 1989.

\bibitem{Singer:99}
P.~Hendriks and M.~Singer.
\newblock Solving difference equations in finite terms.
\newblock {\em J.~Symbolic Comput.}, 27(3):239--259, 1999.

\bibitem{Henn:2013pwa}
J.~Henn.
\newblock {Multiloop integrals in dimensional regularization made simple}.
\newblock {\em Phys. Rev. Lett.}, 110:251601, 2013.

\bibitem{Hoffman}
M.~Hoffman.
\newblock Quasi-shuffle products.
\newblock {\em J. Algebraic Combin.}, 11:49--68, 2000.

\bibitem{GSAGE}
M.~Kauers, M.~Jaroschek, and F.~Johansson.
\newblock Ore polynomials in {S}age.
\newblock In J.~Gutierrez, J.~Schicho, and M.~Weimann, editors, {\em {Computer
  Algebra and Polynomials}}, Lecture Notes in Computer Science, pages 105--125,
  2014.

\bibitem{Kotikov:1990kg}
A.~Kotikov.
\newblock {Differential equations method: New technique for massive {F}eynman
  diagrams calculation}.
\newblock {\em Phys. Lett.}, B254:158--164, 1991.

\bibitem{Laporta:2001dd}
S.~Laporta.
\newblock {High precision calculation of multiloop {F}eynman integrals by
  difference equations}.
\newblock {\em Int. J. Mod. Phys.}, A15:5087--5159, 2000.

\bibitem{Mincer:91}
S.~Larin, F.~Tkachov, and J.~Vermaseren.
\newblock The {FORM} version of {Mincer}.
\newblock Technical Report NIKHEF-H-91-18, NIKHEF, Netherlands, 1991.

\bibitem{Lee:2014ioa}
R.~Lee.
\newblock {Reducing differential equations for multiloop master integrals}.
\newblock {\em JHEP}, 04:108, 2015.

\bibitem{vonManteuffel:2012np}
A.~{\noopsort{Manteuffel}}{van Manteuffel} and C.~Studerus.
\newblock Reduze 2 - distributed {F}eynman integral reduction.
\newblock {\em arXiv: 1201.4330 [hep-ph]}, 2012.

\bibitem{MARSEID}
P.~Marquard and D.~Seidel.
\newblock The {\tt crusher} algorithm.
\newblock unpublished.

\bibitem{Moch:02}
S.~Moch, P.~Uwer, and S.~Weinzierl.
\newblock Nested sums, expansion of transcendental functions, and multiscale
  multiloop integrals.
\newblock {\em J. Math. Phys.}, 6:3363--3386, 2002.

\bibitem{Moch:04}
S.~Moch, J.~Vermaseren, and A.~Vogt.
\newblock The three-loop splitting functions in {QCD}: {T}he non-singlet case.
\newblock {\em Nucl. Phys. B}, 688:101, 2004.

\bibitem{DR3}
E.~Ocansey and C.~Schneider.
\newblock Representing (q-)hypergeometric products and mixed versions in
  difference rings.
\newblock In C.~Schneider and E.~Zima, editors, {\em { Advances in Computer
  Algebra. WWCA 2016.}}, volume 226 of {\em Springer Proceedings in Mathematics
  \& Statistics}, pages 175--213. Springer, 2018.

\bibitem{Petkov:92}
M.~Petkov{\v s}ek.
\newblock Hypergeometric solutions of linear recurrences with polynomial
  coefficients.
\newblock {\em J.~Symbolic Comput.}, 14(2-3):243--264, 1992.

\bibitem{Remiddi:1997ny}
E.~Remiddi.
\newblock {Differential equations for {F}eynman graph amplitudes}.
\newblock {\em Nuovo Cim.}, A110:1435--1452, 1997.

\bibitem{Remiddi:2016gno}
E.~Remiddi and L.~Tancredi.
\newblock Differential equations and dispersion relations for {F}eynman
  amplitudes. the two-loop massive sunrise and the kite integral.
\newblock {\em Nucl. Phys. B}, 907:400--444, 2016.

\bibitem{Remiddi:1999ew}
E.~Remiddi and J.~Vermaseren.
\newblock {Harmonic polylogarithms}.
\newblock {\em Int. J. Mod. Phys.}, A15:725--754, 2000.

\bibitem{Schneider:05a}
C.~Schneider.
\newblock Solving parameterized linear difference equations in terms of
  indefinite nested sums and products.
\newblock {\em J. Differ. Equations Appl.}, 11(9):799--821, 2005.

\bibitem{Schneider:07a}
C.~Schneider.
\newblock Symbolic summation assists combinatorics.
\newblock {\em S\'em.~Lothar. Combin.}, 56:1--36, 2007.
\newblock Article B56b.

\bibitem{Schneider:08c}
C.~Schneider.
\newblock A refined difference field theory for symbolic summation.
\newblock {\em J. Symbolic Comput.}, 43(9):611--644, 2008.

\bibitem{Schneider:15}
C.~Schneider.
\newblock Fast algorithms for refined parameterized telescoping in difference
  fields.
\newblock In M.~W. J.~Guitierrez, J.~Schicho, editor, {\em Computer Algebra and
  Polynomials}, number 8942 in Lecture Notes in Computer Science (LNCS), pages
  157--191. Springer, 2015.


\bibitem{Schneider:13a}
C.~Schneider.
\newblock Simplifying multiple sums in difference fields.
\newblock In C.~Schneider and J.~Bl\"umlein, editors, {\em {Computer Algebra in
  Quantum Field Theory}}, Texts and Monographs in Symbolic Computation, pages
  325--360. Springer, 2013.

\bibitem{Schneider:13b}
C.~Schneider.
\newblock Modern summation methods for loop integrals in quantum field theory:
  The packages {S}igma, {EvaluateMultiSums} and {SumProduction}.
\newblock In {\em {Proc. ACAT 2013}}, volume 523 of {\em J. Phys.: Conf. Ser.},
  pages 1--17, 2014.

\bibitem{DR1}
C.~Schneider.
\newblock A difference ring theory for symbolic summation.
\newblock {\em J. Symb. Comput.}, 72:82--127, 2016.

\bibitem{DR2}
C.~Schneider.
\newblock Summation theory {II}: {C}haracterizations of
  {$R\Pi\Sigma$}-extensions and algorithmic aspects.
\newblock {\em J. Symb. Comput.}, 80(3):616--664, 2017.

\bibitem{DR4}
C.~Schneider.
\newblock Minimal representations and algebraic relations for single nested
  products.
\newblock {\em Programming and Computer Software, in press}, 46(2):in press,
  2020.
\newblock arXiv:1911.04837 [cs.SC].

\bibitem{Steinhauser:2000ry}
M.~Steinhauser.
\newblock {MATAD: A Program package for the computation of MAssive TADpoles}.
\newblock {\em Comput. Phys. Commun.}, 134:335--364, 2001.

\bibitem{Studerus:2009ye}
C.~Studerus.
\newblock {Reduze-{F}eynman Integral Reduction in C++}.
\newblock {\em Comput. Phys. Commun.}, 181:1293--1300, 2010.

\bibitem{vanHoeij:99}
M.~van Hoeij.
\newblock Finite singularities and hypergeometric solutions of linear
  recurrence equations.
\newblock {\em J.~Pure Appl. Algebra}, 139(1-3):109--131, 1999.

\bibitem{Vermaseren:99}
J.~Vermaseren.
\newblock Harmonic sums, {M}ellin transforms and integrals.
\newblock {\em Int. J.~Mod. Phys.}, A14:2037--2976, 1999.

\bibitem{Vermaseren:2005qc}
J.~Vermaseren, A.~Vogt, and S.~Moch.
\newblock The third-order {QCD} corrections to deep-inelastic scattering by
  photon exchange.
\newblock {\em Nucl. Phys. B}, 724:3--182, 2005.

\bibitem{Vogt:2004mw}
A.~Vogt, S.~Moch, and J.~Vermaseren.
\newblock The three-loop splitting functions in {QCD}: {T}he singlet case.
\newblock {\em Nucl.\ Phys.\ B}, 691:129, 2004.
\newblock [hep-ph/0404111].

\bibitem{Zuercher:94}
B.~Z\"urcher.
\newblock {\em Rationale Normalformen von pseudo-linearen Abbildungen}.
\newblock PhD thesis, Mathematik, ETH Z\"urich, 1994.


\end{thebibliography}

\providecommand{\noopsort}[1]{}

\end{document}